# MapReduce Algorithms for Inferring Gene Regulatory Networks from Time-Series Microarray Data Using an Information-Theoretic Approach

Yasser Abduallah, Turki Turki, Kevin Byron, Zongxuan Du, Miguel Cervantes-Cervantes

Jason T. L. Wang

Computer Science Department, New Jersey Institute of Technology, Newark, NJ 07102, USA

## Abstract

Gene regulation is a series of processes that control gene expression and its extent. The connections among genes and their regulatory molecules, usually transcription factors, and a descriptive model of such connections, are known as gene regulatory networks (GRNs). Elucidating GRNs is crucial to understand the inner workings of the cell and the complexity of gene interactions. To date, numerous algorithms have been developed to infer gene regulatory networks. However, as the number of identified genes increases and the complexity of their interactions is uncovered, networks and their regulatory mechanisms become cumbersome to test. Furthermore, prodding through experimental results requires an enormous amount of computation, resulting in slow data processing. Therefore, new approaches are needed to expeditiously analyze copious amounts of experimental data resulting from cellular GRNs. To meet this need, cloud computing is promising as reported in the literature. Here we propose new MapReduce algorithms for inferring gene regulatory networks on a Hadoop cluster in a cloud environment. These algorithms employ an information-theoretic approach to infer GRNs using time-series microarray data. Experimental results show that our MapReduce program is much faster than an existing tool while achieving slightly better prediction accuracy than the existing tool.

# 1. Introduction

Current biotechnology has allowed researchers in various fields to obtain immense amounts of experimental information, ranging from macromolecular sequences, gene expression data to proteomics and metabolomics. In addition to large-scale genomic information obtained through such methods as third generation DNA sequencing, newer technology, such as RNA-seq and ChIP-seq, has allowed researchers to fine tune the analysis of gene expression patterns [1-3]. More information on interactions between transcription factors and DNA, both qualitative and quantitative, is increasingly emerging from microarray data [4-6]. Although microarrays alone do not provide direct evidence of functional connections among genes, the attachment of transcription factors (TFs) and their binding sites (TFBSs), located at specific gene promoters, influences transcription and modulates RNA production from a particular gene, thus establishing a first level of functional interaction. Since the TFs are gene-encoded polypeptides and the target TFBSs belong to different genes, analyses of TFs-TFBSs interactions could reveal gene networks and may even contribute to elucidate unknown GRNs [7]. Besides contributing to infer and understand these interactions, determining GRNs also provides models of such connections [8]. GRNs could be the basis to infer more complex networks, encompassing gene, protein, and metabolic spaces, as well as the entangled and often overlooked signaling pathways that interconnect them [9-13].

The core GRN apparatus consists of the sum of *cis*-regulatory modular DNA sequence elements that interact with TFs. These sequences read and process information incoming from the cell, transducing it into the formation of gene products while modulating their abundance [14]. To make them easier to understand, GRN models must be genome-oriented and viewable at different levels, from global patterns of gene expression, down to *cis*-regulatory DNA and nucleotide sequences [15].

Interactions among genes are mediated by gene products such as DNA-binding proteins (including TFs) and miRNAs. The analyses of gene interactions can be difficult if time-series data are part of the experimental design [16]. Analysis of genes, gene products and metabolism (the Three Spaces of gene networks) would require additional computing resources. Among the previously ignored components of gene networking are miRNAs [17, 18]. In addition to their importance as regulatory elements in gene expression, the capacity of miRNAs to be transported from cell to cell implicates them in a panoply of pathophysiological processes that include antiviral defense, tumorigenesis, lipometabolism and glucose metabolism [16]. This role in disease complicates our understanding of translational regulation via endogenous miRNAs. In addition, miRNAs seem to be present in different types of foods [19] with potential implications on human health and disease. Understanding the biogenesis, transport and mechanisms of action of miRNAs on their target RNA would result in possible therapies, requiring large amounts of computational power, which can be attained by cloud computing and process parallelizing.

Detailed experimental analysis of several functional regulatory elements has revealed that they consist of dense clusters of unique, short DNA sequences specifically recognized by a range of TFs. Biochemically, protein-binding to these sequences controls the regulatory output of the clusters and, from an informational perspective, clustered specific target sites determine the type of regulatory outcome and the cellular functions that will be performed. GRNs are encoded in the DNA and can be thought of as a sequence-dependent regulatory genome, given that TFs recognize specific short DNA motifs. The small length of these motifs means that they will occur frequently but randomly within the enormity of the total genome of a particular organism [20-22]. Therefore, to parse functional regulatory elements using bioinformatics requires the analysis of copious amounts of genomic data.

Analyses of time-series data from microarrays can show the chronological expression of specific genes or groups of genes. These temporal patterns can be used to infer or propose causal relationships in gene regulation [23]. Thus, genes in logical networks can modulate the extent of each other's gene expression over the life span of a cell or a whole organism. Time-series microarray data sheds light on a complex but measurable regulatory system, allowing for a more precise inference of gene interaction.

Numerous algorithms have been developed for inferring GRNs [24-27]. In this paper, we present a new approach, tailored to cloud computing, to infer GRNs using time-series microarray data. Using time as a variable in the analysis of GRNs permits the study of changes in cellular phenotype, as opposed to a snapshot in a limited time frame that may reveal static interactions but not progression of gene expression phenomena. The time-series datasets used in this work come from DREAM4 challenges [28-31] and an Affymetrix Yeast Genome 2.0 Array downloaded from NCBI's Gene Expression Omnibus. The array contains 5,744 probe sets and includes 10,928 *Saccharomyces cerevisiae* genes with 49 time points and transcriptional oscillations of about 4 hours. These oscillations reveal cell redox states, which in turn result from changes in metabolic fluxes and cell cycle phases [32].

As knowledge in several biological fields leads to an ever-expanding accumulation in gene expression data, the main consideration in data processing is that analysis of information becomes increasingly time-consuming, thus creating a demand to speed up the analytical process. In order to obtain results more expeditiously, we develop information-theoretic algorithms using MapReduce that run on a distributed, multi-node Apache Hadoop cluster in a cloud environment. Cloud resources are increasingly more flexible and affordable compared with local traditional computing resources. Cloud computing advantages in the field of bioinformatics research are well known [33-39].

## 2. Materials and Methods

### 2.1. Framework

Previous information-theoretic algorithms for network inference were implemented in the R programming language using steady state data [40] and time-series data [23]. The tool using steady state data is named ARACNE [40] and the tool using time-series data is named TimeDelay-ARACNE [23]. ARACNE infers an undirected network, which basically shows whether two genes are mutually dependent rather than the regulatory relationship between the genes. In contrast, TimeDelay-ARACNE infers a directed network, in which an edge from gene A to gene B indicates that A regulates the expression of B.

In contrast to the R-based ARACNE and TimeDelay-ARACNE, our proposed information-theoretic framework is tailored to the MapReduce programming paradigm. Like TimeDelay-ARACNE [23], the input of our framework is a set of time-series gene expression data and the output is an inferred gene regulatory network. The input dataset contains $m$ genes, and each gene has $n$ expression values recorded at $n$ different time points respectively. Our framework consists of three steps. Step 1 aims to detect, for each gene $g$, the first time point $t$ ($t > 1$) at which a substantial change in the gene expression of *g* with respect to the gene expression of *g* at time point 1 takes place. This *t* is referred to as the time point of Substantial Change of Expression (*ScE*) of gene *g*, denoted $ScE(g)$. Step 2 calculates, for two genes $g_x$ and $g_y$, the influence of $g_x$ on $g_y$, denoted $influence(g_x, g_y)$, based on the *ScE* values of the genes. Step 3 determines the edges between genes using their influence values. Below we present details of the proposed framework.

**Step 1. Calculation of *ScE***

Let $g(t)$ be the expression value of gene *g* at time point *t*. We say *g* is activated (or induced) at time point *t* ($t > 1$) if $\frac{g(t)}{g(1)} > \tau$ where $\tau > 1$ is a threshold. We say *g* is inhibited (or repressed) at time point *t* ($t > 1$) if $\frac{g(t)}{g(1)} < \frac{1}{\tau}$. For each gene *g*, we maintain two sets of time points: $g^+(t)$ and $g^-(t)$; $g^+(t)$ contains all time points at which *g* is induced and $g^-(t)$ contains all time points at which *g* is repressed. Initially, $g^+(t) = \emptyset$ and $g^-(t) = \emptyset$. The two sets of time points are then updated as follows. For each time point *t* ($t > 1$),

$$\text{if } \frac{g(t)}{g(1)} > \tau, \text{ then } g^+(t) = g^+(t) \cup \{t\},$$

$$\text{if } \frac{g(t)}{g(1)} < \frac{1}{\tau}, \text{ then } g^-(t) = g^-(t) \cup \{t\}.$$

If $\frac{1}{\tau} \leq \frac{g(t)}{g(1)} \leq \tau$, then *g* is neither induced nor repressed at time point *t*. In this case, we simply ignore this time point *t* without adding *t* to $g^+(t)$ or $g^-(t)$. The value of $\tau$ used in this study is set to 1.2. With this threshold value and datasets used in the study (DREAM4 [28-31]), there is a significant difference between the mean of the gene expression values of the time points at which *g* is induced and the mean of the gene expression values of the time points at which *g* is repressed according to Student's t-test (p < 0.05).

Let $ScE(g)$ represent the first time point *t* ($t > 1$) at which *g* is either induced or repressed, i.e.,

$$ScE(g) = \min\{g^+(t) \cup g^-(t)\}.$$

For any two genes $g_a$ and $g_b$, there are three cases to be considered.

Case 1: $ScE(g_a) < ScE(g_b)$. We send the ordered pair ($g_a$, $g_b$) and the expression values of the two genes to step 2.

Case 2: $ScE(g_b) < ScE(g_a)$. We send the ordered pair ($g_b$, $g_a$) and the expression values of the two genes to step 2.

Case 3: $ScE(g_a) = ScE(g_b)$. We send $g_a$, $g_b$ with a tag indicating both of the ordered pairs ($g_a$, $g_b$) and ($g_b$, $g_a$) should be considered, together with their gene expression values.

**Step 2. Calculation of influence values**

For each pair of genes ($g_x$, $g_y$) received from step 1, we calculate the time-delayed mutual information [41] between the genes as follows:

$$I^k(g_x, g_y^{(k)}) = \sum_{1 \leq i \leq n-k} p(g_x^i, g_y^{i+k}) \log \frac{p(g_x^i, g_y^{i+k})}{p(g_x^i) p(g_y^{i+k})} \quad (1)$$

where $n$ is the total number of time points, $p(g_x^i)$ is the marginal distribution of $g_x^i$, and $p(g_x^i, g_y^{i+k})$ is the joint distribution of $g_x^i$ and $g_y^{i+k}$. (In our implementation, a hash table is used to calculate the joint distribution to save time and space.) The parameter $k$, $1 \leq k \leq h$, represents the length of delayed time and $h$ is the maximum length of delayed time. (In the study presented here, $h$ is set to 3.) The notation $g_x^i$ denotes the gene expression of $g_x$ at time point $i$ and $g_y^{i+k}$ is the gene expression of $g_y$ at time point $i + k$. Figure 1 illustrates how to calculate time-delayed mutual information. There are 21 time points in Figure 1. The length of delayed time is 2 (i.e., $k$ = 2). Each rectangle represents the gene expression value obtained at some time point. Mutual information of rectangles with the same color is computed. Then, the influence of $g_x$ on $g_y$ is calculated as follows:

$$influence(g_x, g_y) = \max_{1 \leq k \leq h} \{I^k(g_x, g_y^{(k)})\} . \qquad (2)$$

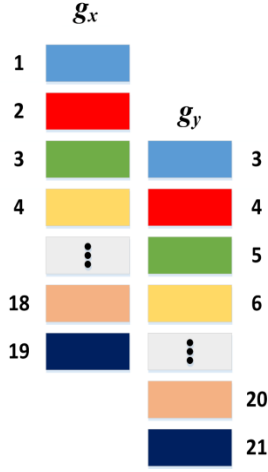

**Figure 1:** Illustration of how to calculate time-delayed mutual information.

Referring to the three cases in step 1, for case 1, we calculate $influence(g_a, g_b)$, and send $(g_a, g_b)$ and $influence(g_a, g_b)$ to step 3. For case 2, we calculate $influence(g_b, g_a)$, and send $(g_b, g_a)$ and $influence(g_b, g_a)$ to step 3. For case 3, if $influence(g_a, g_b) \geq influence(g_b, g_a)$, then we send $(g_a, g_b)$ and $influence(g_a, g_b)$ to step 3; otherwise we send $(g_b, g_a)$ and $influence(g_b, g_a)$ to step 3.

### Step 3. Determination of edges between genes

Let $\varepsilon$ be a threshold. For each pair $(g_x, g_y)$ received from step 2, if $influence(g_x, g_y) > \varepsilon$ then we create an edge from $g_x$ to $g_y$ indicating $g_x$ substantially influences $g_y$ or $g_x$ regulates the expression of $g_y$, i.e., there is a predicted present edge from $g_x$ to $g_y$. If $influence(g_x, g_y) \leq \varepsilon$ then we do not create an edge from $g_x$ to $g_y$, i.e., there is a predicted absent edge from $g_x$ to $g_y$. With the predicted present and absent edges, we are able to infer or reconstruct a gene regulatory network. The value of $\varepsilon$ used in this study is set to 0.96. With this threshold value and datasets used in the study (DREAM4 [28-31]), there is a significant difference between the mean of the influence values of the predicted present edges and the mean of the influence values of the predicted absent edges according to Student's t-test (p < 0.05).

## 2.2. MapReduce Algorithms

Figure 2 presents a high-level conceptual description of the Hadoop MapReduce implementation of our proposed framework. The system includes a driver, and one or more mappers and reducers. The driver takes the input from the user, including the locations of input and output files, as well as algorithm parameters and thresholds. The driver prepares a job with the required configuration, sends the job to Hadoop to start it, and calculates the time the job takes to complete. The mappers are user-defined programs (UDPs), which prepare data and perform calculations, if needed, and then send the processed data (key-value pairs) to the reducers. The reducers are also UDPs, which perform the final processing and write the results into the output file. Hadoop optimizes the number of mappers for a job. The user can control the number of reducers needed for completing the job. The Hadoop distributed file system (HDFS) is a global repository for storage of the input flat file (in plain text format) with gene expression data and the output file with an inferred gene regulatory network.

Each gene has an identifier. (We use $g_x$ to represent both a gene and its identifier when the context is clear.) Each line in the input file contains a pair of genes and their expression values. Genes are sorted based on their identifiers. Each pair of genes $g_x$, $g_y$ occurs in the input file only once; specifically, the gene pair in which the identifier of $g_x$ is less than the identifier of $g_y$ occurs in the input file. Suppose there are $m$ genes. There are $(m \times m-1)/2$ lines in the input file.

We develop four MapReduce algorithms, named M0, M1, M2, and M3 respectively. These algorithms differ in which steps, as described in Section 2.1, are performed by mappers.

**Algorithm M0.** In this algorithm, mappers perform zero steps. Reducers have to do steps 1, 2 and 3. In the key-value pairs transmitted between mappers and reducers, the key is a pair of genes, and the value contains the expression profiles of the genes.

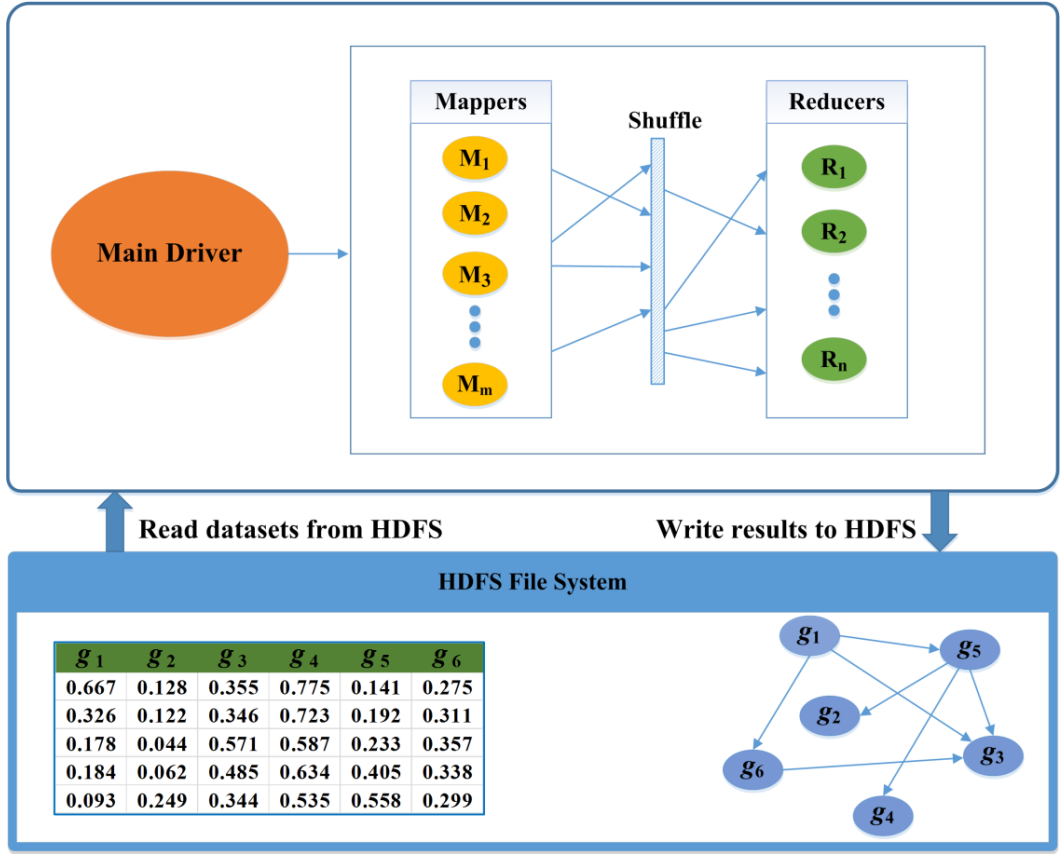

**Figure 2:** Conceptual description of the Hadoop MapReduce implementation of our proposed algorithms.

**Algorithm M1.** In this algorithm, mappers perform step 1. Reducers have to do steps 2 and 3. In the key-value pairs transmitted between mappers and reducers, the key is a pair of genes and the value contains the expression profiles of the genes. Part of the value is a tag indicating which case in step 1 applies to the pair of genes.

**Algorithm M2.** In this algorithm, mappers have to do steps 1 and 2. Reducers perform step 3. In the key-value pairs transmitted between mappers and reducers, the key is an ordered pair of genes ($g_x, g_y$) and the value is $influence(g_x, g_y)$.

**Algorithm M3.** In this algorithm, mappers have to do steps 1, 2 and 3. Reducers perform zero steps. In the key-value pairs transmitted between mappers and reducers, the key is the edge $g_x \rightarrow g_y$ and the value is the influence of $g_x$ on $g_y$ that exceeds the threshold $\varepsilon$.

The time needed by mappers is bounded by $O(m^2/M)$ and the time needed by reducers is bounded by $O(m^2/R)$, where $m$ is the number of genes, $M$ is the number of mappers and $R$ is the number of reducers. Thus, the time complexity of our MapReduce algorithms is $O(m^2/M) + O(m^2/R)$. Note that this is a very pessimistic upper bound since reducers often work in parallel with mappers, and hence the actual time needed by the algorithms is much less. Note also that, in practice, $M > R$, and hence the time complexity of our algorithms is bounded by $O(m^2/R)$.

## 3. Results and Discussion

### 3.1. Experimental Results

The four algorithms described in Section 2.2 were implemented in MapReduce and Java on a Hadoop infrastructure (cloud), which is a virtual environment based on VMware Big Data Extensions (BDE). The infrastructure hardware cluster associated with BDE is comprised of two IBM iDataPlex dx360 M4 servers. Each dx360 M4 server is comprised of two Intel Xeon 2.7GHz E5-2680 (8 Core) CPUs for a total of 16 cores per server. With the enabling of hyperthreading, the number of logical processors is doubled to provide 32 logical processors per server. Each server has 128 GB RAM.

The dataset used in the experiments was GSE30052 [32], downloaded from the Gene Expression Omnibus (GEO) at http://www.ncbi.nlm.nih.gov/geo/. This dataset was created using an Affymetrix Yeast Genome 2.0 Array containing 5,744 probe sets for *S. cerevisiae* gene expression analysis. The dataset contains 10,928 genes with 49 time points. The dataset is split into key-value pairs as described in Section 2.2 and the input file has (10,928 × 10,927)/2 lines, taking up 26.8 GB of disk space. Hadoop assigns 254 mappers to this dataset. The default value for the number of reducers is set to 20.

We divided GSE30052 into smaller datasets that were subsets of GSE30052 with varying numbers of genes. Figure 3 compares the running times of the four MapReduce algorithms described in Section 2.2 for varying dataset sizes. It can be seen from the figure that all the four algorithms scale well when the dataset size becomes large (i.e., the number of genes increases). Algorithm M2 performs the best. This happens because with M2, the mappers, working in parallel, share some workload with the reducers, which perform a relatively smaller amount of computation while writing results into the output file. It is worth noting that M2 is better than M3, in which the mappers have to do all the computation. Algorithm M0 performs the worst. With M0, the reducers have to do all the calculations and become too busy to quickly complete the job.

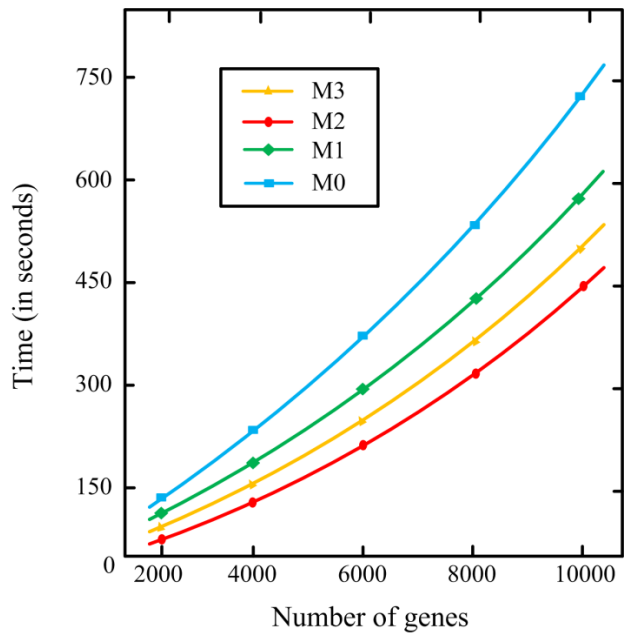

**Figure 3:** Performance comparison of the four MapReduce algorithms.

We then fixed the algorithm and used M2 in all subsequent experiments. Figure 4 shows running times of the M2 algorithm with 1, 20, and 100 reducers respectively. It can be seen that the optimal number of reducers is 20. With this configuration, the reducers work at their maximum limit. When there are too many (e.g., 100) reducers, the overhead is too large, and as a consequence the system is slowed down. On the other hand, when only one reducer is employed, the reducer is overloaded and the overall performance of the system degrades.

Finally, we conducted experiments to compare the MapReduce implementation of the M2 algorithm running on the Hadoop cluster (denoted MRC), the MapReduce implementation of the M2 algorithm running on a standalone single-node server (denoted MRS), the Java implementation of the M2 algorithm running on a single-node server, and the R implementation of the related time-delayed mutual information algorithm, TimeDelay-ARACNE [23]. In Figure 5, it can be seen that MRC is highly scalable and that it outperforms the other three programs. Notably, due to Hadoop's overhead, MRS is even slower than the Java program. The R program is not scalable; its running time dramatically increases as the dataset becomes large.

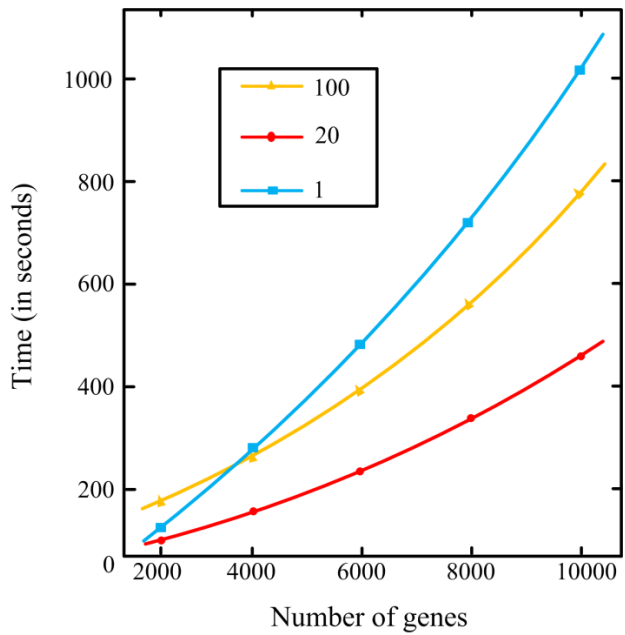

**Figure 4:** Effect of the number of reducers on the performance of the M2 algorithm.

## 3.2. Discussion

Our information-theoretic algorithms for inferring gene regulatory networks are implemented in MapReduce and run on a Hadoop cluster. A tool that is closely related to our work is the TimeDelay-ARACNE program in R [23], which also infers gene regulatory networks from time-series gene expression profiles using an information-theoretic approach. As shown in Figure 5, the TimeDelay-ARACNE program in R does not scale well whereas our MapReduce program is highly scalable when running on the Hadoop cluster. Furthermore, our MapReduce program differs from the TimeDelay-ARACNE R program in that our algorithm is deterministic whereas the R program is implemented based on a non-deterministic algorithm, specifically Markov random fields. For the same dataset and parameter values, the R program produces different results in different executions. In contrast, our MapReduce program always produces

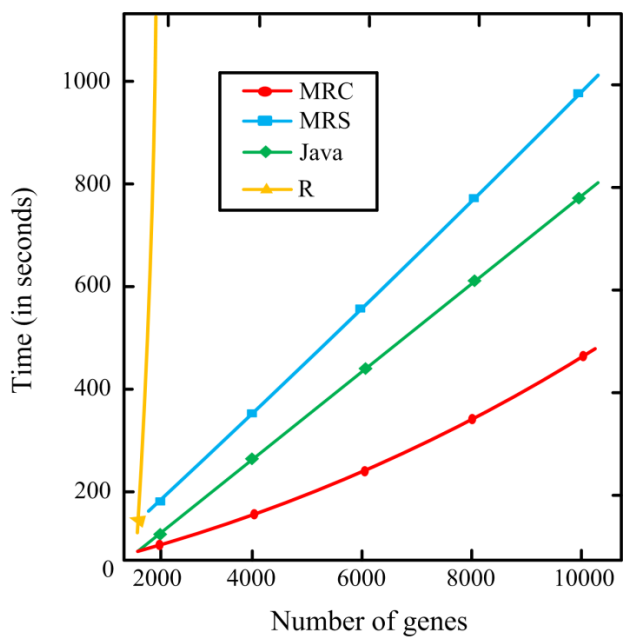

**Figure 5:** Performance comparison of the MapReduce algorithm (M2) and related programs.

the same result, in different executions, for the same dataset.

To evaluate the accuracy of these programs, we adopted the five time-series gene expression datasets available in the DREAM4 100-gene in silico network inference challenge [28-31]. Each dataset contains 10 times series, where each time series has 21 time points, for 100 genes. Thus, in each time series, each gene has 21 gene expression values. Totally, there are 50 time series in the five datasets. Each time-series dataset is associated with a gold standard file, where the gold standard represents the ground truth of the network structure for the time-series data. Each edge in the gold standard represents a true regulatory relationship between two genes. Our experimental results showed that the average accuracy of TimeDelay-ARACNE on the five datasets is approximately 92.4%. The average accuracy of our deterministic algorithm (M2) is about 93.1%, which is slightly better than TimeDelay-ARACNE. It was observed that, in a dataset, when M2 is better than TimeDelay-ARACNE, it has a higher accuracy than TimeDelay-ARACNE for every execution of TimeDelay-ARACNE on the dataset. Furthermore, our MapReduce program (M2) takes, on average, 20 seconds to infer a gene regulatory network on a DREAM4 dataset while the average time used by the TimeDelay-ARACNE R program is 3,500 seconds.

We also tested on different values for the parameters $\tau$, $\varepsilon$, and the maximum length of delayed time, $h$, used in the proposed algorithms. Experimental results showed that the default values for these parameters ($\tau=1.2$, $\varepsilon=0.96$, $h=3$) achieve the highest accuracy. When compared with other parameter values (e.g., the maximum length of delayed time $h=6$, $\varepsilon=0.35$ or $\tau=2$), the accuracy achieved by the default parameter values is significantly higher than the accuracy achieved by the other parameter values according to Wilcoxon signed rank tests [42] ($p < 0.05$).

## 4. Conclusions

We have presented four MapReduce algorithms for reconstructing gene regulatory networks from time-series microarray data using an information-theoretic approach. Our experimental results showed that the algorithm (M2) that uses mappers to perform a large portion of work and reducers to perform a relatively small amount of computation achieves the best performance. This M2 algorithm is faster than an algorithm in which the mappers have to do all the computation. Moreover, the M2 algorithm is much faster than another algorithm in which the reducers have to do all the computation and become too busy to quickly complete the job.

When tested on DREAM4 datasets with 100 genes in each dataset, our MapReduce program (M2) is slightly better than a closely related R program (TimeDelay-ARACNE [23]) in terms of accuracy; furthermore, our MapReduce program is much faster than the existing R program. When tested on a big dataset (GSE30052 [32]) with 10,928 genes, our MapReduce program was found to be highly scalable whereas the R program was not (cf. Figure 5). It should be pointed out, however, that the comparison with the TimeDelay-ARACNE R program is not completely fair. Our MapReduce program is based on a parallel algorithm whereas the TimeDelay-ARACNE R program is based on a sequential algorithm. Further study would be needed to investigate parallel versions of the R program or a new TimeDelay-ARACNE R package that supports parallelization.

The work presented here shows that distributing highly parallel tasks in a cloud environment achieves higher performance than running the tasks in a standalone or non-cloud environments. In general, cloud computing can provide the power to integrate the ever-increasing information about the Three Spaces of gene networks [8] as well as the multi-pronged signal transduction pathways traversing these spaces. Comprehending systems biology and functional genomics could eventually contribute to a better grasp of organismal physiology. Thus, the cloud would provide computing power that is needed as the analysis of multilevel processes becomes more complicated. Cloud computing will enable genome-scale network inference as demonstrated in this study.

Epigenetics [43] is an emerging aspect of gene regulation whose study would require enormous computing capacity. This type of posttranslational regulation cross-talk involves chemical modifications of DNA and histones in a process known as chromatin remodeling. The role of genetic and epigenetic networks in a variety of health conditions is now coming into view. For example, there are at least 450 different genes associated with intellectual disability and related cognitive disorders. Some of these genes are involved in synaptic plasticity and cell signaling whereas others are epigenetic genes involved in chromatin modifications [44]. Analysis of the interactions across these genes and networks, as well as finding new mutations, will require the development of highly expeditious bioinformatics tools to mine the anticipated high amounts of data.

Genome-scale metabolic models are becoming essential in biomedical applications, and researchers are moving towards building such models [45]. MapReduce algorithms could become a powerful tool in the analyses of all aspects of gene networking in the Three Spaces paradigm. In general, cloud computing could facilitate the handling of the vast amounts of information (big data) that such analyses require.

## Acknowledgments

The authors thank N. Patel and L. Zhong for useful conversations concerning gene network inference.